\documentclass[%
 reprint,
 amsmath,amssymb,
 aps,
]{revtex4-2}
\usepackage[T1]{fontenc}
\usepackage{graphicx}% Include figure files
\usepackage{dcolumn}% Align table columns on decimal point
\usepackage{bm}% bold math
\usepackage[shortlabels]{enumitem}

\begin{document}
\preprint{APS/123-QED}

\title{(111) Si spin qubits constructed on L point of band structure}% 

\author{Takafumi Tokunaga}
\affiliation{Department of Physics, Waseda University, Tokyo 169-8555, Japan}
\author{Hiromichi Nakazato}
\affiliation{Department of Physics, Waseda University, Tokyo 169-8555, Japan}

\date{\today}% It is always \today, today,

\begin{abstract}

\text{(001)} Si spin qubits are being intensively studied because they have structures similar to that of CMOS devices currently being produced, and thus have the advantage of utilizing state-of-the-art miniaturization, integration, and variation-reduction-technologies. However, there are still issues, such as further improvement of relaxation and decoherence time, stabilization of valley-splitting control, and reduction of the variation caused by the roughness of the interface. In this study, new measures are proposed to address these three issues. Instead of confining an electron to the minimum energy point \(\text{X}_0\) of the conduction band along the band structure \(\Gamma\)-\(\Delta\)-X in \text{(001)} Si crystals, we propose confining an electron to the L point along \(\Gamma\)-\(\Lambda\)-L in (111) Si crystals. At the \(\text{X}_0\) point, the symmetry causes spin-orbit interaction to act on the electron, and the sixfold degeneracy is lifted into a fourfold and a twofold, and the valley-splitting of the twofold conflicts with the two-level system. In the symmetry of the L point, substantial spin polarization disappears, facilitating the reduction of the spin-orbit interaction, and the fourfold degeneracy is lifted to threefold and single, and the single becomes ground state. The need to increase the magnitude of the valley-splitting is exempt, allowing the electron to be controlled away from the interface, which is expected to reduce the variation caused by the roughness of the interface. Data on the confinement of an electron to the L point and the control of fourfold degeneracy are needed, and it is hoped that prototype silicon spin qubits constructed on (111) Si crystals will be developed and that the proposed device structure will help implement quantum computers based on Si devices.

\begin{description}
\item[DOI:]
XXXXXXXX

\end{description}
\end{abstract}

\maketitle

%\tableofcontents
\section{\label{sec:1}INTRODUCTION\protect \lowercase{} }

There is a growing activity to develop spin qubit devices for quantum computers using Si semiconductor fabrication techniques \cite{kawakami2014electrical,veldhorst2014addressable,veldhorst2015two,burkard2023semiconductor,watson2018programmable,zajac2018resonantly}. The reasons for this are as follows.

(1). In nature, Si materials are 92.23\verb|%| nuclear spin-free \({ }^{28} \mathrm{Si}\). Therefore, by increasing this purity, hyperfine coupling due to nuclear spins can be reduced \cite{itoh1999growth,yoneda2018quantum,kato2003host}.

(2). Spin-orbit coupling is relatively small in Si crystals because Si is a light element with the atomic number 14, and Si has a diamond structure that has inversion symmetry \cite{elliott1954theory,zwanenburg2013silicon}.

(3). When developing a spin qubit device with a structure similar to that of CMOS devices currently being produced, the prototyping of qubit devices can proceed in existing development and production lines \cite{zwerver2022qubits,neyens2024probing}.

(4). In addition to miniaturization and integration technologies, variation-reduction-technologies have been established as mass production technologies for CMOS devices, and these can be directly applied to the development of spin qubit devices \cite{asai2023device,cifuentes2024bounds}.

In the development of qubit devices for quantum computers, the biggest development challenge is to reduce relaxation and decoherence in quantum states. When an external magnetic field is applied to a spin qubit, the energy is separated into two levels according to the direction of its spin by Zeeman separation. When this two-level system is used as a single qubit, the main causes of relaxation and decoherence are as follows \cite{hanson2007spins}.

\begin{enumerate}[(a)]
    \item  The hyperfine coupling due to nuclear spins \cite{witzel2007decoherence}
    \item  The spin-orbit coupling in crystals \cite{winkler2003spin}
    \item  The spin-orbit coupling due to interfaces \cite{winkler2003spin,cifuentes2024bounds}
\end{enumerate}

Countermeasures against (a) are being promoted by using epitaxially grown \({ }^{28} \mathrm{Si}\) crystals by CVD gas using high purity \({ }^{28} \mathrm{Si}\). In fact, \({ }^{28} \mathrm{Si}\) epitaxial layers with a purity of 99.992\verb |%|  or higher have been achieved, and it has been confirmed that the concentration of \({ }^{29} \mathrm{Si}\) is less than 0.006\verb |%|, and the concentration of \({ }^{30} \mathrm{Si}\) is less than 0.002\verb |%| \cite{mazzocchi201999}.

Using high purity crystals \({ }^{28} \mathrm{Si}\), it has already been possible to achieve relaxation time \(T_1\) and decoherence time \(T_2\) of several hundred milliseconds or more \cite{tyryshkin2012electron,maurand2016cmos}. Further improvements in \(T_1\) and \(T_2\) are being pursued through further purification. Therefore, the major issue is to orient the measures against (b) and (c) and to proceed with the orientations. This study mainly examines countermeasures against (b) and (c). 

The spin-orbit coupling for an atom with atomic number \(Z\) is known to be approximately proportional to the fourth power of \(Z\) \cite{slater1961quantum}. Therefore, spin-orbit coupling in Si with atomic number 14 is only 3.7\verb|%| of that in Ge (atomic number 32) with the same crystal symmetry, however even this small spin-orbit coupling can be a source of relaxation and decoherence in Si.

The approach taken in this study to overcome (b) is to change the Si crystal used from conventional \text{(001)} Si to (111) Si. Details on the reasons for this and the expected benefits are discussed in Subsection \(\text {II-A}\).

In the band structure of Si crystals, the energy minimum point \(\text{X}_0\) of the conduction band exists in the \text{(001)} direction in the \(k\) space \cite{feher1959electron}. To take advantage of the symmetry in the (111) direction, a new method to confine an electron to the L valley is considered necessary. Specific methods are discussed in Subsection \(\text {II-B}\).

Also, at the L point, the wave function of the electron has a fourfold degenerate valley structure, as in the case of Ge \cite{chazalviel1975spin}, therefore it is necessary to take measures to prevent the lifting of this degeneracy from interfering with the control of the qubit as a two-level system. The situation of valley-splitting at the point L is discussed in Subsection \(\text {II-C}\).

The measures against (c) are to stably suppress the roughness at the \(\mathrm{SiO}_2\)/Si or SiGe/Si interface to the level of an atomic step and to confine the electron away from the interface so that its wave function does not contact the interface. Specific countermeasures are described in Subsection \(\text {II-D}\).

In this paper, we exclusively consider thoroughly reducing the spin-orbit interaction. We will discuss the usage of the spin-orbit interaction to electrically manipulate the qubits in a forthcoming paper.

Conclusions are provided in Section \(\text {III}\).

\section{Results and Discussions}
\subsection{Reducing the spin-orbit coupling in Si crystals}

\begin{figure}
    \centering
    \includegraphics[width=0.85\linewidth]{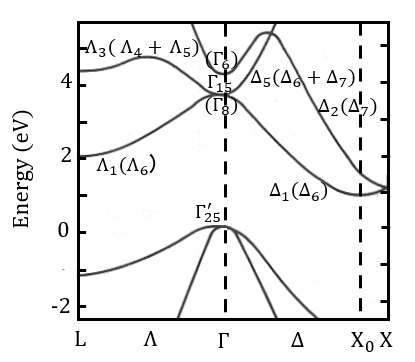}
    \caption{The band structure of Si near the \(\Gamma\) point. The symmetry representations without brackets are expressed by single group. The symmetry expressions with brackets are according to double group taking spin into account.}
    \label{fig:1}
\end{figure}

Si crystals have a diamond structure and their symmetry is characterized in terms of bulk space group and site point group. The symmetry of the Si crystal with the quantum well is noncentrosymmetric, belonging to the \(T_d(\overline{4} 3 \mathrm{~m})\) bulk space group \cite{winkler2003spin}. In nonmagnetic and noncentrosymmetric
crystals, the Dresselhaus spin-orbit interaction and the Rashba spin-orbit interaction act to split the spin-up and spin-down energy bands in the crystal. In each split band, spin polarization is observed. 

The conditions under which spin-orbit coupling occurs depend on the direction of the crystal growth axis along which the quantum well is formed. The reason for this is that the site point group for each crystal axis direction corresponds to \(C_{4 v}(4 \mathrm{mm})\) for the \text{(001)} direction, \(C_{2 v}(\mathrm{mm} 2)\) for the \text{(011)} direction, and \(C_{3 v}(3 \mathrm{m})\) for the (111) direction. In this paper, the direction of the crystal growth axis is taken to be the \(z\)-direction, and the directions perpendicular to the crystal growth axis are taken to be the \(x\) and \(y\)-directions.

\text{Figure 1} shows the band structure of Si near the \(\Gamma\) point. We focus on the lowest-energy band and the second-lowest-energy band of the conduction band. In the tight-binding picture, the lowest-energy band, \(s\)-like (\(l = 0\)) or \(p\)-like (\(l = 1\)) states are called anti-bonding \(s\)-like conduction band states or anti-bonding \(p\)-like conduction band states. In the case of Si, the anti-bonding \(p\)-like \(\Gamma_{15}\) band has the lowest energy in the conduction band. 

When spin degrees of freedom are included and spin-orbit coupling is taken into account, the \(\Gamma_{15}\) band splits into the \(j = 3 / 2\) (\(l = 1\), \(\text{S} = 1/2\)) state of the \(\Gamma_{8}\) band and the \(j = 1 / 2\) (\(l = 1\), \(\text{S} = 1/2\)) state of the \(\Gamma_{6}\) band.
Since the spin symbols are confusing with the \(s\) orbitals, they are indicated by arrows whenever possible, and capital S is used when necessary. Along the \(\Gamma\)-\(\Delta\)-X axis, the \(\Gamma_{8}\) band is further separated into the \(\Delta_{6}\) band and the \(\Delta_{6}+\Delta_{7}\) band (both two-dimensional \(p\)-like state) \cite{dresselhaus1967fourier,elliott1954spin}.

On the other hand, along the \(\Gamma\)-\(\Lambda\)-L axis, the \(\Gamma_{8}\) band splits into a doubly degenerate band \(\Lambda_{6}\) (two-dimensional \(s\)-like state) and two overlapping non-degenerate bands \(\Lambda_{4}+\Lambda_{5}\) (each one-dimensional \(p\)-like state). Next, spin-orbit interactions for the two symmetry axes \(\Gamma\)-\(\Delta\)-X and \(\Gamma\)-\(\Lambda\)-L are compared in perspective of the effects of multipoles on the band structure \cite{dresselhaus1967fourier,elliott1954spin}.

\subsubsection{The effects of multipoles on the band structure from the microscopic viewpoint}

The Hamiltonian of one electron on the band structure is expressed as follows.

\begin{eqnarray}
\begin{aligned}
{H}=\sum_{X}^{Q,M,T,G} \sum_{\boldsymbol{k} \sigma \sigma^{\prime}} \sum_{l m} {X}_{l m}^{\mathrm{ext}} {X}_{l m}^{\sigma \sigma^{\prime}}(\boldsymbol{k}) c_{\boldsymbol{k} \boldsymbol{\sigma}}^{\dagger} c_{\boldsymbol{k}  \boldsymbol{\sigma}^{\prime}},
\end{aligned}
\end{eqnarray}
where \(c_{\boldsymbol{k} \boldsymbol{\sigma}}^{\dagger}\left(c_{\boldsymbol{k} \boldsymbol{\sigma}}\right)\) is the creation (annihilation) operator of an electron with the wave vector \(\boldsymbol{k}\) and  spin \(\boldsymbol{\sigma}\), and \({X}_{l m}\) are the multipoles with the azimuthal and magnetic quantum numbers, \(l\) and \(m\). It is known that there are four types of multipoles according to their spatial inversion and time reversal properties: electric (\({Q}\): poler/true tensor), magnetic (\({M}\): axial/pseudotensor), magnetic toroidal (\({T}\): polar/true tensor), and electric toroidal (\({G}\): axial/pseudotensor) multipoles. \({X}_{l m}^{\mathrm{ext}}\) represent symmetry-breaking fields of which the microscopic origins are the crystalline electric field, the external field applied to the system, and external strain field, and so on \cite{hayami2018classification,hayami2018microscopic}. Writing from the low-rank multipole contribution, the Hamiltonian can be expressed as

\begin{eqnarray}
\begin{aligned}
{H}= & \sum_{\boldsymbol{k} \sigma \sigma^{\prime}}\left[\frac{\hbar^2 \boldsymbol{k}^2}{2 m} \sigma_0+\boldsymbol{Q}^{\mathrm{ext}} \cdot(\boldsymbol{k} \times \boldsymbol{\sigma})+\boldsymbol{M}^{\mathrm{ext}} \cdot \boldsymbol{\sigma}\right. \\
& \left.+\boldsymbol{T}^{\mathrm{ext}} \cdot \boldsymbol{k} \sigma_0+G_0^{\mathrm{ext}}(\boldsymbol{k} \cdot \boldsymbol{\sigma})+\cdots\right]_{\sigma \sigma^{\prime}} c_{\boldsymbol{k} \sigma}^{\dagger} c_{\boldsymbol{k} \sigma^{\prime}}.
\end{aligned}
\end{eqnarray}

In the first term, the form \(Q_0^{\mathrm{ext}}=\hbar^2 / 2 m\) is taken, and the second term represents the spin-orbit interaction, and here we focus on the second term. \(\boldsymbol{Q}^{\mathrm{ext}}\) represent the symmetry-breaking field of which the microscopic origins are the crystalline electric field and the external electric field applied to the qubit. For example, \(\boldsymbol{M}^{\mathrm{ext}}\), \(\boldsymbol{T}^{\mathrm{ext}}\), and \(G_0^{\mathrm{ext}}\) arise from an external magnetic field, electric current, and rotation, respectively \cite{hayami2018classification}.

The band structure is characterized by the breaking of the spatial inversion symmetry which causes the spin-orbit interaction. The Hamiltonian of Rashba spin-orbit interaction
is expressed as

\begin{eqnarray}
\begin{aligned}
H_{R S O}= & \boldsymbol{Q}^{e x t} \cdot \boldsymbol{Q}=\boldsymbol{Q}^{e x t} \cdot(\boldsymbol{k} \times \boldsymbol{\sigma}) \\
= & Q_x^{e x t}\left(k_y \sigma_z-k_z \sigma_y\right)+Q_y^{e x t}\left(k_z \sigma_x-k_x \sigma_z\right) \\
& +Q_z^{e x t}\left(k_x \sigma_y-k_y \sigma_x\right),
\end{aligned}
\end{eqnarray}
where \(Q_x=k_y \sigma_z-k_z \sigma_y, \ Q_y=k_z \sigma_x-k_x \sigma_z, \ Q_z=k_x \sigma_y-k_y \sigma_x,\) and \(\boldsymbol{Q}^{e x t}\) is the same electrical polar vector as the electric field and the odd rank E multipoles (\(\ell \geq 1\)) \cite{hayami2018classification}. The Hamiltonian of Dresselhaus spin-orbit interaction is expressed as

\begin{eqnarray}
\begin{aligned}
H_{D S O} & =Q_{x y z}^{e x t} Q_{x y z}=\sqrt{15} Q_{x y z}^{e x t}\left[k_x\left(k_{y}^2-k_z^2\right) \sigma_x\right. \\
& \left.+k_y\left(k_z^2-k_x^2\right) \sigma_{y}+k_z\left(k_x^2-k_y^2\right) \sigma_z\right], 
\end{aligned}
\end{eqnarray}
where \(Q_{x y z}=\sqrt{15}\left(k_y k_z Q_x+k_z k_x Q_y+k_x k_y Q_z\right)\), and \(Q^{e x t}_{x y z}\) corresponds to electric octupole (\(\ell \geq 3\)) \cite{hayami2018classification}. In the case of Si, the contribution of Dresellhausa spin-orbit interaction is considered to be very small, since the \(s\) and \(p\) orbitals are dominant and there is negligible influence of the orbitals with rank greater than three (\(f, h, \cdots\)). Even when there is a contribution from Dresellhausa spin-orbit interaction, \text{Eq. (4)} shows that in the (111) direction, \(H_{D S O} = 0\) due to \(k_x=k_y=k_z\), which is more advantageous than in the \text{(001)} direction.

In the following, electron orbitals are assumed to be \(s\) and \(p\) orbitals, and the basis set of the wave functions, which also takes into account the spin degrees of freedom, is treated as  \(|s\uparrow\rangle\), \(|s\downarrow\rangle\), \(\left|p_x \uparrow\right\rangle\), \(\left|p_x \downarrow\right\rangle\), \(\left|p_y \uparrow\right\rangle\), \(\left|p_y \downarrow\right\rangle\), \(\left|p_z \uparrow\right\rangle\), and \(\left|p_z \downarrow\right\rangle\).

\subsubsection{Comparison of \(\Delta_{1}(\Delta_{6})\) and  \(\Lambda_{1}(\Lambda_{6})\)}

The minimum energy point \(\text{X}_0\) of the conduction band of Si is  \(k_0=0.85 \times {2 \pi}/{a}\) in the \text{(001)} direction of \(k\)-space (\({a}\) is the Si lattice constant 5.43 \AA). On the other hand, the L point of the conduction band is at the edge of the Brillouin zone in the (111) direction, which has a higher symmetry than the \(\text{X}_0\) point \cite{hsueh2024engineering}. The \(\text{X}_0\) point has relatively low crystal symmetry, and it is intuitively strange why there is an energy minimum in the conduction band at this point, the cause of which has recently begun to be investigated \cite{oliphant2024does}. 

E. Oliphant \(et \ al\). discuss how the band structure is caused by the interaction of bonding, antibonding, and nonbonding orbitals in a 3D crystal structure, taking Si crystals as an example. According to them, along the \(\Gamma\)-\(\Delta\)-X axis, \(\Delta_{1}\) is a \(p_z\)-like state, and the fraction of \(p_z\) orbitals is evaluated to be more than \(80 \%\) and that of \(s\) orbitals is less than \(20 \%\). They find that the dip near the X point originates from a cosine shape along \(\Gamma\)-X  arising from the second nearest neighbor \(p_z\)-\(p_z\) bonding, combined with a positive linear slope due to the first nearest neighbor \(s\)-\(s\), \(s\)-\(p_z\), and \(p_z\)-\(p_z\) interactions \cite{oliphant2024does}.

Along the \(\Gamma\)-\(\Lambda\)-L axis, \(\Lambda_{1}\) is an \(s\)-like state, and the fraction of \(s\) orbitals is evaluated to be \(55 \%\), and the fraction of \(p_z\) orbitals is \(45 \%\) \cite{oliphant2024does}. Consequently, the \(\Delta_{1}\) band along the \(\Gamma\)-\(\Delta\)-X axis is the \(p_z\)-like state and spin-orbit interaction acts on it. However, \(\Lambda_{1}\) is an \(s\)-like state and the fraction of \(p_z\) orbitals is approximately \(35 \%\) lower than that of the \(\Delta_{1}\) band; therefore, the spin-orbit interaction along the \(\Lambda_{1}\) band is considered to be smaller than that along the \(\Delta_{1}\) band.

In addition to the above, we note that the ground state at the L point has the following characteristics which will be discussed in II-C. That is, the ground state at the L point after valley-splitting has an iso-energetic surface of a rotational ellipsoid in the radial direction along the \(z\) axis, as shown in \text{Fig. 2}. The parts of the crystal structure that form the hexagonal structure shown are in fact composed of two triangular structures that are obtained by inversion operations on each other. The L point is located at the center of the hexagonal structure. The centers of these two triangular structures are separated from each other by 1/4 of the lattice constant in real space. The two triangular structures have \(p_z\) dipole moments in opposite directions, and these \(p_z\) dipole moments cancel each other. Therefore, the spin-orbit interaction of the entire hexagonal structure vanishes. This concept corresponds to the so-called hidden spin polarization concept \cite{zhang2014hidden}.

\subsubsection{Comparison of \(\Delta_{2}(\Delta_{7})\) and  \(\Lambda_{3}(\Lambda_{4}+\Lambda_{5})\)}

The \(\Delta_{5}\) band is also a \(p_z\)-like state, but its energy increases rapidly as it approaches X, therefore it is out of the category of the lowest and second-lowest bands in the conduction band. The second lowest energy band near X is the \(\Delta_{2}\) band. In the \(\Delta_{2}\) band, the fraction of \(p_z\) orbitals increases as it approaches X, and conversely, the fraction of \(s\) orbitals increases as it approaches \(\Gamma\). The \(\Delta_{2}\) band is a \(p_z\)-like state near the \(\text{X}_0\) point, similar to the \(\Delta_{1}\) band \cite{oliphant2024does}, and spin-orbit interaction acts on it.

On the other hand, the situation is more complicated in the \(\Lambda_{3}\) band, where the two bands overlap. This is because a unique situation occurs in site symmetry \(C_{3 v}(3 \mathrm{m})\) in the (111) direction, which is explained below.

Kai Liu \(et \ al\). discuss the phenomenon of spin polarization appearing and vanishing depending on the site point group for each crystal axis direction in GaAs with the bulk space group \(T_d(\overline{4} 3 \mathrm{m})\) \cite{liu2019band}. Along the \text{(001)} direction of the crystal axis, that is, along the \(\Gamma\)-\(\Delta\)-X axis, as usual, the energy band splits due to spin-orbit coupling and shows spin-polarization. 

On the other hand, along the (111) direction, i.e., along the \(\Gamma\)-\(\Lambda\)-L axis, the energy band splits as well, but surprisingly, they conclude that pure spin-polarization disappears in both split bands and call the phenomenon band splitting with vanishing spin polarizations (BSVSP) \cite{liu2019band}.

\begin{table}
\caption{
Character Table of the Double Group of  \(\Lambda.\) \(\Lambda_1, \Lambda_2, \Lambda_3\) are the single groups. \(\Lambda_4, \Lambda_5, \Lambda_6\) are the double groups with spin degrees of freedom. In the Fig.1, \(\Lambda_1\) is transformed to \(\Lambda_6\), and \(\Lambda_3\) is transformed to \(\Lambda_4+\Lambda_5\) \cite{elliott1954spin}.
}
\label{tab:1}%
\begin{ruledtabular}
\begin{tabular}{ccccccc}
 &${E} $&$\bar{E}$
 &$2 C_3$&$2 \bar{C}_3$&$3I \times C_2$&$3I \times \bar{C}_2$\\
\hline
\(\Lambda_1\) & 1 & 1 & 1 & 1
& 1 & 1 \\
\(\Lambda_2\) & 1 & 1 & 1 & 1
& -1 & -1 \\
\(\Lambda_3\) & 2 & 2 & -1 & -1
& 0 & 0 \\
\(\Lambda_4\ (\Lambda_3 \times D_{1/2}\)) & 1 & -1 & -1 & 1
& \textit{i} & -\textit{i} \\
\(\Lambda_5\ (\Lambda_3 \times D_{1/2}\)) & 1 & -1 & -1 & 1
& -\textit{i} & \textit{i} \\
\(\Lambda_6\ (\Lambda_{1,2,3} \times D_{1/2}\)) & 2 & -2 & 1 & -1
&  0 & 0 \\
\end{tabular}
\end{ruledtabular}
\end{table}

We apply the mechanism of BSVSP generation to Si, which has the same crystal symmetry \(T_d\). The site symmetry along the \(\Gamma\)-\(\Lambda\)-L axis belongs to \(C_{3 v}(3 \mathrm{m})\), and the character table of the double group which takes spin degrees of freedom into account is shown in Table I \cite{elliott1954spin}. Along the \(\Gamma\)-\(\Lambda\)-L axis, the \(\Gamma_{8}\) band splits into two overlapping non-degenerate bands \(\Lambda_{4}+\Lambda_{5}\) (each one-dimensional \(p\)-like state) and a doubly degenerate band \(\Lambda_{6}\).

As can be seen from the character table, two overlapping non-degenerate bands \(\Lambda_{4},\Lambda_{5}\) are independent of each other and are connected by a time-reversal operation. Therefore, they have spin polarizations in opposite directions, and when these two bands are superimposed, the spin polarization disappears, which is called BSVSP \cite{liu2019band}.

As mentioned above, the spin-orbit interaction works as usual in the \(\Delta_{2}\) band of the \text{(001)} direction. On the other hand, in the \(\Lambda_{3}\) band of the (111) direction, it is found to have a special symmetry \(C_{3 v} \otimes D_{{1}/{2}}\) in which the spin-orbit interaction does not work in effect due to the BSVSP phenomenon. 

\begin{figure}
    \centering
    \includegraphics[width=0.90\linewidth]{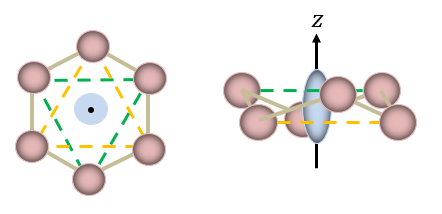}
    \caption{The hexagonal structure at the L point. The hexagonal structure consists of two triangular structures (located in inversion symmetry with respect to each other), each with opposite sign of \(p_z\) dipole field in the \(z\)-direction. These \(p_z\) dipole moments cancel each other, therefor the spin-orbit interaction of the entire hexagonal structure vanishes.}
    \label{fig:2}
\end{figure}

The reason why the BSVSP phenomenon occurs in the (111) direction is investigated below from the microscopic perspective of the Si crystal structure. The Hamiltonian of the electron along the \(\Lambda\) axis is expressed as follows.
\begin{eqnarray}
\begin{aligned}
H_{\Lambda}= & H_0+H_{R S O}+H_{D S O}, \\
\end{aligned}
\end{eqnarray}
where \(H_0\) is the spin-independent part, \(H_{R S O}\) is the Rashba spin-orbit interaction part \text{Eq. (3)}, and \(H_{D S O}\) is the Dresselhaus spin-orbit interaction part \text{Eq. (4)}.
The wave functions of the \(p\) orbitals, which have opposite spin directions can be denoted by \(\left|p_z, \text{S}_z\right\rangle\) ,\(\left|p_z, -\text{S}_z\right\rangle\) e.g. in the \(z\)-direction.
Calculating the expectation value of energy due to these \(p\) orbitals  \(\left|p_\alpha, \text{S}_z\right\rangle\) and \(\left|p_\alpha, -\text{S}_z\right\rangle\), taking into account that \(k_x=k_y=k_z\) along the \(\Lambda\) axis, yields the following equation
\begin{eqnarray}
\begin{aligned}
& \sum_{\alpha=x, y, z}\left\langle p_\alpha, \mathrm{S}_z\right| H_{\Lambda}\left|p_\alpha, \mathrm{S}_z\right\rangle \\
& \quad=\sum_{\alpha=x, y, z}\left\langle p_\alpha,-\mathrm{S}_z\right| H_{\Lambda}\left|p_\alpha,-\mathrm{S}_z\right\rangle .
\end{aligned}
\end{eqnarray}
This equation indicates that two \(p\) orbitals, with spin in opposite directions to each other, have the same energy, i.e., they overlap in the same band. 

\begin{figure}
    \centering
    \includegraphics[width=0.75\linewidth]{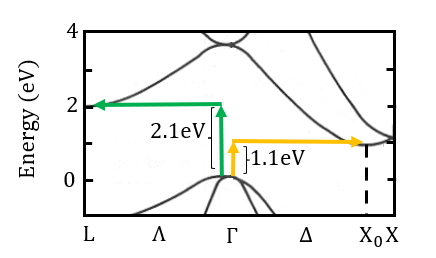}
    \caption{The indirect optical transitions to the \(\text{X}_0\) point (yellow arrow) and the L point (green arrow) from the point of maximum energy in the valence band. The L point is forming a valley similar to that of the \(\text{X}_0\) point. Both optical transitions have been experimentally confirmed.}
    \label{fig:3}
\end{figure}

\subsection{\label{sec:2}Confining an electron to the L point of the band structure}
As discussed in \(\text {II-A}\), to utilize the symmetry features along the \(\Gamma\)-\(\Lambda\)-L axis and to reduce spin-orbit coupling, it is necessary to confine an electron to the L point rather than to the \(\text{X}_0\) point. The minimum energy point \(\text{X}_0\) of the conduction band of Si is in the \text{(001)} direction \(k_0=0.85 \times {2 \pi}/{a}\) of \(k\)-space (\({a}\) is the Si lattice constant 5.43 \AA). On the other hand, the L point of the conduction band is in the (111) direction, which has a higher symmetry than the \(\text{X}_0\) point \cite{hsueh2024engineering}. However, as already mentioned, the energy minimum of the conduction band is not at the L point. 

If the maximum energy in the valence band is set to 0 eV as the energy reference, the \(\text{X}_0\) point is at a height of approximately 1.1 eV and the L point is at a height of approximately 2.1 eV as shown in Fig. 3. The difference between the \(\text{X}_0\) and L points is 1.0 eV. That is, the L point has an energy 1.0 eV higher than the \(\text{X}_0\) point.

Point L has a downward energy slope at all \(\Gamma \rightarrow\) L and \(\mathrm{K} \rightarrow\) L points, forming a valley similar to that of point \(\text{X}_0\). The optical transitions \(\Gamma\rightarrow \mathrm{X}_0\) and \(\Gamma \rightarrow \mathrm{L}\) have been found to be possible according to indirect optical transitions, as shown in \text{Fig. 3}, and their absorption spectra have been experimentally confirmed \cite{chelikowsky1976nonlocal}.

\begin{figure}
    \centering
    \includegraphics[width=0.75\linewidth]{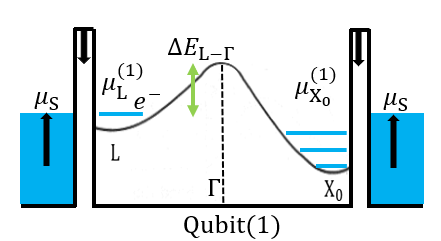}
    \caption{The process of how an electron is confined to the L point. First, the chemical potential \(\mu_{\mathrm{S}}\) is gradually increased, filling electrons from the ground level at the \(\text{X}_0\) point of Qubit(1) and stopping when the level at the L point is filled with one electron. In real space, the source is located on one side and the drain on the opposite side, but in this figure, the sources are placed on both sides to make the positional relationship to the \(\text{X}_0\) point and the L point equal.}
    \label{fig:4}
\end{figure}

\begin{figure}
    \centering
    \includegraphics[width=0.75\linewidth]{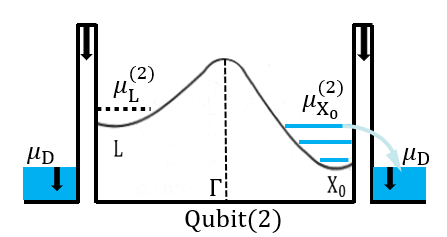}
    \caption{The process of how the electrons are drained to the drain. In real space, the source is located on one side and the drain on the opposite side, but in this figure, the sources are placed on both sides to make the positional relationship to the \(\text{X}_0\) point and the L point equal.}
    \label{fig:5}
\end{figure}

\begin{figure}
    \centering
    \includegraphics[width=1.00\linewidth]{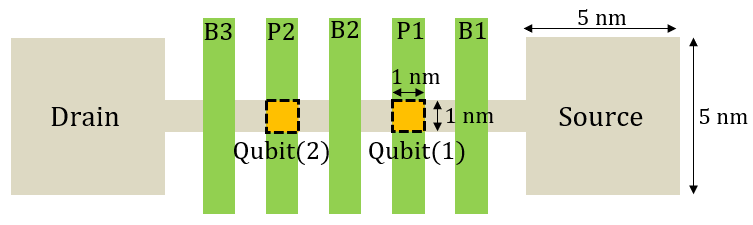}
    \caption{Layout of the qubits to confine an electron to the L point, with two identical qubits lined up, the source on one side and the drain on the opposite side. Qubit(1) and Qubit(2) are assumed to be identical cubic quantum dots with a side length of 1 nm, and the source and drain are each assumed to be a cubic quantum dot with a side length of 5 nm.}
    \label{fig:6}
\end{figure}

Next, we investigate the specific process of how the electron is confined to the L point.  As shown below, in practice, the source is formed on one side of the quantum dot and the drain is formed on the opposite side, but the source and drain positions are shown as if they were formed on both sides so that the relative positions of the source and drain are equivalent to the \(\text{X}_0\) point and the L point in Fig. 4 and Fig. 5. The lowest energy band of the conduction band shown in the figure is not the entire band, but only a part of it, near the \(\text{X}_0\) point and the L point.

We first consider the number of electrons that can be stored in the band with the lowest energy in the conduction band. Expressing the wave number vector \(\boldsymbol{k}\) as a first-order combination of reciprocal lattice vectors \(\boldsymbol{b_1}\), \(\boldsymbol{b_2}\), \(\boldsymbol{b_3}\),

\begin{figure}
    \centering
    \includegraphics[width=0.75\linewidth]{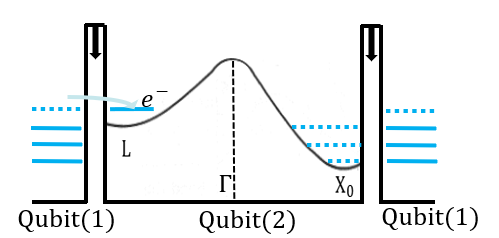}
    \caption{The process of how an electron is confined to the L point. When barrier gate B2 is controlled to lower the barrier between Qubit(1) and Qubit(2), the electron injected into the L point of Qubit(1) tunnels to Qubit(2).}
    \label{fig:7}
\end{figure}

\begin{eqnarray}
\begin{aligned}
& \boldsymbol{k}=\frac{n_1}{N_1} \boldsymbol{b}_1+\frac{n_2}{N_2} \boldsymbol{b}_2+\frac{n_3}{N_3} \boldsymbol{b}_3 \\
& n_i=0,1, \cdots, N_i-1,\ (i=1,2,3),\\
\end{aligned}
\end{eqnarray}
where the numbers \(N_1, N_2\), and \(N_3\) are integers. The number of electron levels formed in one band is \(N=N_1 N_2 N_3\), which depends on the size of the quantum dot in which the qubit is formed, and the number of electrons that can be stored is \(2N\), taking into account the spin degrees of freedom. Therefore, assuming that the quantum dot is a cube, we determine the number of electrons that can be stored in the band with the lowest conduction band energy, using \(N\) as the number of Si unit cells when the length of one side of the cube is varied. The results are summarized in the Table \(\text{II}\).

\begin{table}
\caption{
Estimate of the number of electrons that can be occupied in the lowest energy band of the conduction band.
}
\label{tab:table1}%
\begin{ruledtabular}
\begin{tabular}{ccccc}
 \(\text{Length of one side}\) &${10 \text{ nm}}$ &${5 \text{ nm}}$ &${2 \text{ nm}}$ &${1 \text{ nm}}$\\
\hline
\(\text{Number of unit cells (\(N\))}\) & 6250 & 781 & 50 & 6\\
\(\text{Number of energy levels}\) & 6250 & 781 & 50 & 6\\
\(\text{Number of electrons}\) & 12500 & 1562 & 100 & 12\\
\end{tabular}
\end{ruledtabular}
\end{table}

The method of confining an electron to the L point is explained using a layout in which two identical qubits are arranged between the source and the drain as shown in \text{Fig. 6}. In this layout of the qubits, we prepare \raise0.2ex\hbox{\textcircled{\scriptsize{1}}} the source and drain made of cubes of \((5 \mathrm{~nm})^3\), and \raise0.2ex\hbox{\textcircled{\scriptsize{2}}} the quantum dots that form Qubit(1) and Qubit(2) made of cubes of \((1 \mathrm{~nm})^3\). 

In the source and drain, the numbers of levels in the lowest energy band are sufficiently large as shown in Table II, and the energy distributions are considered to be continuous. On the other hand, in the Qubits(1) and (2), the numbers of levels in the lowest energy band are 6, so the energy distributions are considered to be discontinuous. If we assume that the fourth level corresponds to the level at the L point, the electrons are packed from the ground state, and the seventh electron is injected into the L point.

Let the energy difference between the L and \(\Gamma\) points be \(\Delta E_{\mathrm{L}-\Gamma}\) as shown in \text{Fig. 4}. The electrochemical potentials of the source and drain are indicated by \(\mu_{\mathrm{S}}\) and \(\mu_{\mathrm{D}}\) respectively, and the electrochemical potentials of the \(\text{X}_0\) point and the L point of Qubit\((i)\) are indicated by \(\mu^{(i)}_{\text{X}_0}\) and \(\mu^{(i)}_L\), respectively. 

By lowering the barrier between the source and Qubit (1) controlling the barrier gate B1 and gradually increasing \(\mu_{\mathrm{S}}\), the electrons are filled one by one from the lower energy levels at the \(\text{X}_0\) point, and when \(\mu_{\mathrm{S}}\) is increased to \(\mu_L^{(1)} \leq \mu_{\mathrm{S}}<\mu_L^{(1)}+\Delta E_{\mathrm{L}-\Gamma}\), the seventh electron becomes the first electron to be injected into the L point, based on the assumption we made earlier as shown in Fig. 4. The number of electrons injected can be confirmed by the generation of current. The injection of electrons is stopped here. 

Next, as shown in \text{Fig. 7}, we consider shuttling the electron injected into the L point of Qubit(1) to the L point of the adjacent Qubit(2) by means of tunneling. When the barrier gate B2 is controlled to lower the barrier between Qubits(1) and (2), the electron injected into the L point of Qubit(1) tunnels to Qubit(2). In this case, Qubits(1) and (2) are designed identically, and based on the conservation law of the energy and momentum of the electron, it is considered that there is a high probability that the electron moves to the L point of Qubit(2). It is also possible that the electron moves to the \(\text{X}_0\) point of Qubit(2), accompanied by relaxation phenomena such as phonon emission, but it is not possible to predict the probability of this tunneling with the relaxation process.

If the probability of an electron being injected into the L point is close to 1, then when a series of injections are carried out, in a situation where one electron already exists at the L point in Qubit(2), electron transfer does not occur due to the Coulomb blockade phenomenon. On the other hand, if the probability that an electron is injected into the \(\text{X}_0\) point in Qubit(2) is sufficiently high, then six electrons are injected in succession, as assumed. If the injection of electrons is monitored by measuring the current value, it is possible to determine whether the electrons are being injected into the L point or the \(\text{X}_0\) point. 

If the magnitude of the current at the time of injection reveals that it has been injected into the L point, the initialization is complete, and the qubit state can be transferred directly to the quantum calculation process. If it is found that the injection point is \(\text{X}_0\), the chemical potential of the drain is set to \(\mu_D<\mu^{(2)}_{\text{X}_0}\), and all the electrons injected into the \(\text{X}_0\) point are drained as shown in \text{Fig. 5} and the injection process is carried out again. At present, the probability of injection into the L point cannot be predicted because the injection path and barrier height to the L point within the crystal are not predictable. We plan to investigate the injection method for the L point with a high probability, including reducing the relaxation to the \(\text{X}_0\) point.

It is hoped that spin qubit devices constructed on (111) Si crystals will be developed and experimental data will be obtained to increase the probability of injections into the L point.

\subsection{Controlling the valley-splitting at the L point}

If an electron is confined into the \(\text{X}_0\) point, there are six equivalent valleys in the Brillouin zone, therefore, the electron state is sixfold degenerate and has an iso-energetic surface of a long rotational ellipsoid in the radial direction as shown in Fig. 8 (a). The six valleys are denoted by \(\text {A, B, C, D, E, \text{and}\ F}\) in Fig. 8 (b). In quantum dots with a crystal growth axis in the \text{(001)} direction, the applied voltage for the confinement in the \(z\)-direction causes valley-splitting, whereby the twofold (\(z\)-direction: \(\text{A, B}\)) and fourfold (\(x,y\) directions: \(\text{C, D, E, F}\)) degenerate states. The effective masses at the conduction band minimum are anisotropic, with the longitudinal effective mass more than four times larger than the transverse. This difference in effective masses gives the twofold degenerate ground state and the fourfold degenerate exited state. The splitting is further enhanced if tensile stress is applied in the \(z\)-direction, whereby the twofold degeneracy is lifted \cite{ando1982electronic,friesen2010theory,saraiva2011intervalley}. Consequently, the ground state is the lower energy state from the twofold degeneracy. It is essential to prevent this valley-splitting from interfering with the control of the qubit as a two-level system.

\begin{figure}
    \centering
    \includegraphics[width=0.85\linewidth]{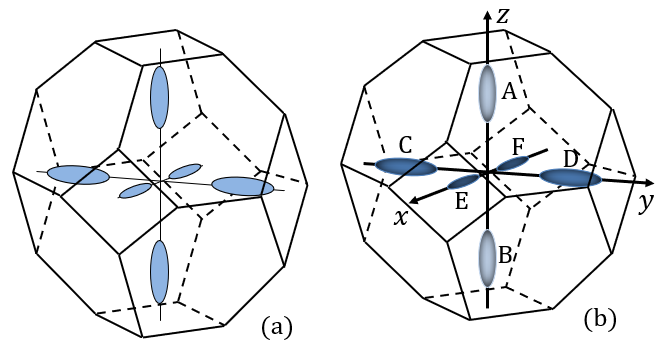}
    \caption{(a) The six-fold degenerate iso-energetic surface at the \(\text{X}_0\) point. (b) The applied voltage for the confinement in the \(z\) direction causes valley-splitting, whereby the twofold (\(z\)-direction: \(\text{A, B}\)) and fourfold (\(x-y\) direction: \(\text{C, D, E, F}\)) degenerate states.}
    \label{fig:8}
\end{figure}

\begin{figure}
    \centering
    \includegraphics[width=0.50\linewidth]{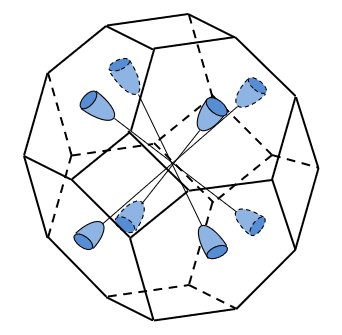}
    \caption{The four-fold degenerate iso-energetic surface at the L point. The iso-energetic surfaces of a rotational ellipsoid are split in half because they are on the edge of the Brillouin zone.}
    \label{fig:9}
\end{figure}

On the other hand, if the electron is confined into the L point, we predict that valley-splitting will occur as follows. There are four equivalent L points in the Brillouin zone, i.e., the electron state is fourfold degenerate and has an iso-energetic surface of a rotational ellipsoid in the radial direction as shown in Fig. 9. In quantum dots with a crystal growth axis in the (111) direction, the applied voltage for the confinement in the \(z\)-direction causes valley-splitting, whereby the single (\(z\)-direction: \(\text{A}\)) and threefold (\(\text{B, C, D}\)) degenerate states as shown in Fig. 10. The valley \(\text{A}\) presents its major axis and a high effective mass to the \(z\)-direction, while the other three have their major axes in directions other than the \(z\)-direction and present a lower effective mass. This difference of effective masses gives the single ground state and the threefold degenerate exited state. Consequently, the ground state is the single state even if the tensile stress is further applied in the \(z\)-direction. As mentioned in \(\text {II-A}\), the L point is in the center of the hexagonal structure, and the four hexagonal structures form a tetrahedron with the center being the \(\Gamma\) point. The single ground state is located in the center of the hexagonal structure as shown in Fig. 10. 

These predictions are supported by an analogy from valley-splitting in Ge, which has the same crystal symmetry and where the L point is the minimum energy point of the conduction bands. In Ge, it has been theoretically and experimentally confirmed that single (\(z\)-direction) and triple valley-splitting with the single state being the ground state exist \cite{reuszer1964optical,li2012intrinsic}.

\begin{figure}
    \centering
    \includegraphics[width=0.75\linewidth]{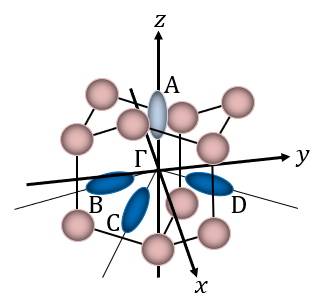}
    \caption{The single (\(z\)-direction) and triple valley-splitting. The L point is at the center of the hexagonal structure, and the four hexagonal structures form a tetrahedron which center is the \(\Gamma\) point. The single (\(z\)-direction: \(\text{A}\)) and triple valley-splitting (\(\text{B, C, D}\)) are caused by applying voltage or stress to the quantum dot in the \(z\)-direction, and non-degenerate single state (\(\text{A}\)) becomes the ground state.}
    \label{fig:10}
\end{figure}

For Si spin qubits constructed on the L point of the band structure in the (111) direction, a single ground state would be very advantageous in the operation of the device. In a qubit device with a crystal growth axis in the direction \text{(001)}, a study has been conducted to ensure that the lifting of the double degeneracy does not conflict with the control of the qubit. To ensure that the higher energy state in the two splitting states does not conflict with the two-level system, an attempt is being made to increase valley-splitting by confining the electron close to the interface and utilising their spin-orbit interactions. 

In a qubit device with a crystal growth axis in the (111) direction and a quantum well in the \(z\)-direction, confining electrons into the L point is exempted from the requirement for confining the electron close to the interface.

\subsection{Reducing the spin-orbit coupling caused by the interface}
Spin-orbit coupling has been found to increase due to interfacial roughness in MOS structures (\(\text {Metal / SiO}_2   / \text{ Si}\)) or heterostructures (e.g. \(\text {SiGe / Si / SiGe}\)) \cite{cifuentes2024bounds,pena2023utilizing}. When a voltage is applied to the electrode to confine the electron and draw it closer to the interface just below the electrode, spin-orbit coupling due to the roughness of the interface becomes large and its variation increases. 

\begin{figure}
    \centering
    \includegraphics[width=0.75\linewidth]{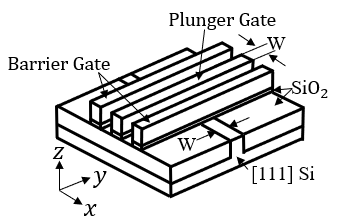}
    \caption{The qubit device with the planarization of the fin. The plunger gate controls the potential of the quantum dots under the electrode to confine an electron. Barrier gates control the tunneling of the electron between quantum dots, injecting the electron from the source, ejecting the electron to the drain, etc.}
    \label{fig:11}
\end{figure}

\begin{figure}
    \centering
    \includegraphics[width=0.75\linewidth]{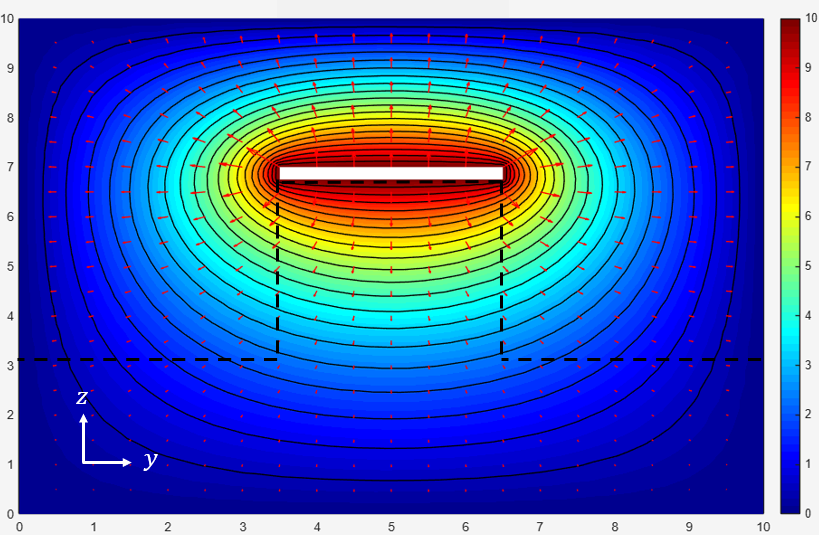}
    \caption{The potential distribution in the \(z-y\) plane. The dotted line shows the contour of (111) Si, and the white area is  \(\text {SiO}_2\) part. The boundary condition of the Poisson's equation at the interface between  \(\text {SiO}_2\) and Si is given as a Dirichlet boundary condition and simulated by MATLAB package. The magnitude of the potential is divided into 10 levels and displayed in color.}
    \label{fig:12}
\end{figure}

Therefore, it is necessary to reduce the roughness of the interfaces in stacked structures (\(\text{SiO}_2 \ / \ \text{Si} \) or \(\text{SiGe} \  / \ \text{Si}\)) as a countermeasure. In the heterostructure \(\text { SiGe / Si / SiGe }\), specific measures to reduce the roughness of the interface, such as the adjustment of the atomic ratio of Si and Ge, are in progress and can be expected to have some effect \cite{scappucci2021crystalline,paquelet2022atomic}. 

\begin{figure}
    \centering
    \includegraphics[width=0.75\linewidth]{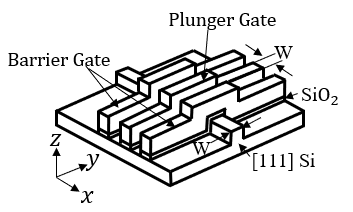}
    \caption{The qubit device with the protruding fin. The plunger gate controls the potential of the quantum dots under the electrode to confine an electron. Barrier gates control the tunneling of the electron between quantum dots, injecting the electron from the source, ejecting the electron to the drain, etc.}
    \label{fig:13}
\end{figure}

\begin{figure}
    \centering
    \includegraphics[width=0.75\linewidth]{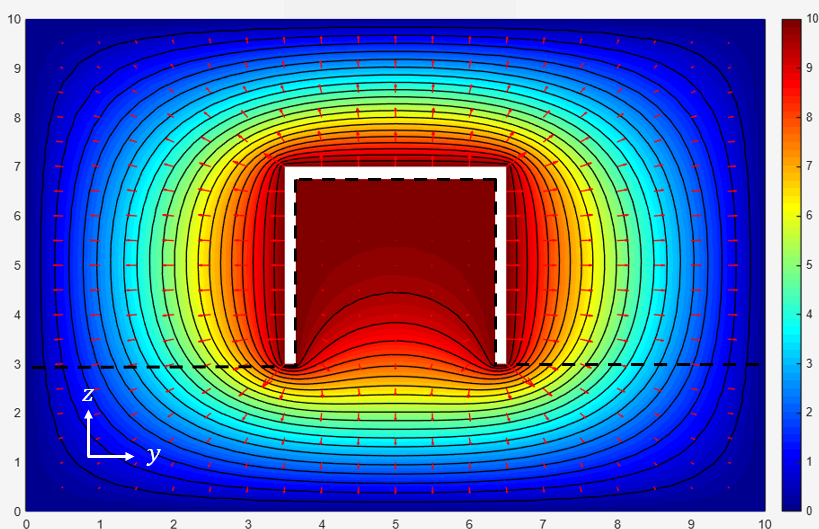}
    \caption{The potential distribution in the \(z-y\) plane. The dotted line shows the contour of (111) Si, and the white area is  \(\text {SiO}_2\) part. The boundary condition of the Poisson's equation at the interface between  \(\text {SiO}_2\) and Si is given as a Dirichlet boundary condition and simulated by MATLAB package. The magnitude of the potential is divided into 10 levels and displayed in color.}
    \label{fig:14}
\end{figure}

In the MOS structure, \(\text {SiO}_2\) is amorphous and the roughness of the \(\text{SiO}_2 \ / \ \text{Si} \) interface cannot be completely eliminated, so it is necessary to consider confinement methods that prevent the electron wave function from reaching the interface.

Figure 11 shows a qubit device with a fin-based MOS structure. The substrate is (111) crystal Si and when \({ }^{28} \mathrm{Si}\) is epitaxially grown on the substrate, the crystal growth axis direction is aligned in the same (111) direction. The fins have a width of W and are planarised by an oxide film formed to cover the sides of the fins. The \(x\)-direction is defined as the direction of the fin. The width of the electrodes is W, the same as that of the fins, and they intersect perpendicularly with the fins on the planarised plane. By making the widths of the fins and electrodes equal, voltage can be applied from the electrodes in the \(z\)-direction to induce a symmetrical potential in the \(x\) - \(y\) plane within the fins, which is considered advantageous for reducing spin-orbit coupling \cite{madhav1994electronic,avetisyan2012strong}.

The potentials induced in this qubit are simulated when voltage is applied to the electrodes. The result of a potential distribution in the \(z\) - \(y\) plane is shown in Fig. 12. The potential gradient near the \(\text{SiO}_2 \ / \ \text{Si} \) interface just below the electrode is sharp, and electrons are easily drawn to the interface. 

As another option, Fig. 13 shows a qubit device with a fin-based MOS structure, but the electrodes are shaped to straddle the protruding fins without being flattened by an oxide film. Also in this case, by making the fins and electrodes equal in width, voltage can be applied from the electrodes in the \(z\)-direction to induce a symmetrical potential in the \(x\) - \(y\) plane within the fins, which is considered advantageous for reducing spin-orbit coupling \cite{madhav1994electronic,avetisyan2012strong}.

In this qubit, we simulate the potential induced when a voltage is applied to the electrodes for electron confinement. The result of a potential distribution in the \(z\) - \(y\) plane is shown in \text{Fig. 14}, and it is found that the potential gradient in the \(z\)-direction within the fins is relatively flat. In this potential distribution with a small \(z\)-directional gradient in the fins, an electron is confined away from the \(\text{SiO}_2 \ / \ \text{Si} \) interface, and spin-orbit interaction due to the interface is considered to be prevented. In this case, the potential gradient is still in the \(z\)-direction, that is, in the (111) direction; therefore, the suppression of the spin-orbit interaction, as described in \(\text {II-A}\), can be expected.

In Subsection \(\text{II-B}\), the number of Si unit cells is calculated from the volume of the quantum dot forming the qubit, assuming that it is a cube. From this number of Si unit cells, the number of electron levels that can exist in the lowest energy band of the conduction band is calculated, so the validity of the assumption that the quantum dot is a cube is extremely important. As can be seen in \text{Fig. 14}, in this structure, the real quantum dot can be approximated as a cube, confirming the validity of the previous assumption.

\section{Conclusions}

When Si spin qubit devices are fabricated on a \text{(001)} Si substrate and coherent states are created by confining the electron to the \(\text{X}_0\) valley of the conduction band, spin-orbit interaction acts on the electron, causing relaxation and decoherence.

On the other hand, if the device is fabricated on a (111) Si substrate and the electron can be confined to the L valley of the conduction band, the spin-orbit coupling does not act on the electron due to the site symmetry feature and relaxation and decoherence can be reduced. 

The confinement of electrons to the L valley can be achieved by raising the electrochemical potential of the sources made of highly doped Si above that of the L point. The amount of current flowing between the quantum dot and sources confirms whether confinement to the L point is successful or not.

For the electron confined to the L valley, applying a voltage in the \(z\)-direction causes the fourfold degeneracy to be lifted into a single state and a threefold degenerate state. Since the ground state is the single state, there is no obstacle as a two-level system due to valley-splitting, and there is no need to consider increasing the amount of valley splitting by using the interface roughness.

This makes it possible to keep the electron apart from the interface, thereby reducing factors of variation such as the spin-orbit interaction caused by the interface roughness. In order to keep the electron apart from the interface, we propose a structure in which the electrodes straddle the protruding fins without flattening the fins with an oxide film in a MOS structure qubit device.

\bibliographystyle{ieeetr}
\bibliography{references}

\end{document}